Refractive-index-sensing radio-frequency comb with intracavity multi-mode interference fibre sensor

Ryo Oe[1,2,†], Shuji Taue[3,†], Takeo Minamikawa[2,4], Kosuke Nagai[1], Yasuhiro Mizutani[2,5], Tetsuo Iwata[2,4], Hirotsugu Yamamoto[2,6], Hideki Fukano[3], Yoshiaki Nakajima[2,7], Kaoru Minoshima[2,7], and Takeshi Yasui[2,4,*]

[1]Graduate School of Advanced Technology and Science, Tokushima University, 2-1, Minami-Josanjima, Tokushima, Tokushima 770-8506, Japan

[2]JST, ERATO, MINOSHIMA Intelligent Optical Synthesizer Project, 2-1, Minami-Josanjima, Tokushima, Tokushima 770-8506, Japan

[3]Graduate School of Natural Science and Technology, Okayama University, 3-1-1 Tsushima-naka, Kita-ku, Okayama, Okayama 700-8530, Japan

[4]Graduate School of Technology, Industrial and Social Sciences, Tokushima University, 2-1, Minami-Josanjima, Tokushima, Tokushima 770-8506, Japan

[5]Graduate School of Engineering, Osaka University, 2-1, Yamadaoka, Suita, Osaka 565-0871, Japan

[6]Center for Optical Research and Education, Utsunomiya University, 7-1-2, Yoto, Utsunomiya, Tochigi 321-8585, Japan

[7]Graduate School of Informatics and Engineering, The University of Electro-Communications, 1-5-1 Chofugaoka, Chofu, Tokyo 182-8585, Japan

†Co-first author (Equal contribution to the first author)

*Corresponding authors' e-mail address: yasui.takeshi@tokushima-u.ac.jp

**Abstract**

Optical frequency combs have attracted attention as optical frequency rulers due to their tooth-like discrete spectra together with their inherent mode-locking nature and phase-locking control to a frequency standard. Based on this concept, their applications until now have been demonstrated in the fields of optical frequency metrology and optical distance metrology. However, if the utility of optical combs can be further expanded beyond their optical-frequency-ruler-based application by exploiting new aspects of optical combs, this will lead to new developments in optical metrology and instrumentation. Here, we report a fibre sensing application of optical combs based on a coherent frequency link between the optical and radio-frequency regions, enabling high-precision refractive index measurement of a liquid sample based on frequency measurement in radio-frequency region. Our technique encodes a refractive index change of a liquid sample into a radio-frequency comb by a combination of an intracavity multi-mode-interference fibre sensor and wavelength dispersion of a cavity fibre. Then, the change in refractive index is read out by measuring the repetition frequency of the radio-frequency comb with a frequency counter and a frequency standard. Use of an optical comb as a photonic radio-frequency converter will lead to the development of new applications in high-precision fibre sensing with the help of functional fibre sensors and precise radio-frequency measurement.

# 1. Introduction

An optical frequency comb (OFC) [1-3] is regarded as a group of a vast number of phase-locked narrow-linewidth continuous-wave (CW) light sources with a constant frequency spacing $f_{rep}$ (typically, 50–100 MHz) over a broad spectral range. The inherent mode-locking nature and active laser control enable us to use an OFC as an optical frequency ruler traceable to a microwave or radio-frequency (RF) frequency standard. Based on the concept of an optical frequency ruler, OFCs have found several applications in optical frequency metrology and distance metrology; examples include atomic spectroscopy [2], gas spectroscopy [4], solid spectroscopy [5], spectroscopic ellipsometry [6], strain sensing [7], and distance measurement [8]. If the application fields of OFCs could be expanded beyond their optical-frequency-ruler-based application, the adoption of OFCs as next-generation lasers in optical metrology and instrumentation will grow.

One interesting feature of OFCs is a coherent frequency link between the optical and RF regions. For example, when an OFC is detected with a photodiode, its quadratic-detection function converts the OFC into a secondary frequency comb in the RF region without changing $f_{rep}$; that is to say, an RF comb is generated. While the frequency uncertainty of the OFC is transferred into that of the RF comb via such a coherent detection process, use of an RF comb simplifies the experimental methodology because measurement in the RF region benefits from high precision, high functionality, convenience, and low cost by making use of various kinds of RF measurement apparatuses. Therefore, RF combs have been applied to optical distance metrology, such as long-distance measurement with extremely high precision [9-11].

The key feature required to enable this new use of RF combs is an RF conversion function in a fibre OFC cavity. The repetition frequency $f_{rep}$ in a ring cavity is given by

$$f_{rep} = \frac{c}{nL}, \quad (1)$$

where $c$ is the speed of light in vacuum, $n$ is the refractive index of the cavity fibre, and $L$ is the geometrical length of the fibre cavity. If an external disturbance, such as a temperature change, strain, or vibration, interacts with the fibre OFC cavity, $f_{rep}$ shifts as a result of a change of the optical cavity length $nL$. In other words, the OFC can act as a photonic RF converter for such an external disturbance. Since RF frequency measurement can be performed with high precision and a wide dynamic range by making use of accurate frequency standards, the combination of a photonic RF converter and RF frequency measurement, namely, a photonic RF sensor, has the potential to greatly enhance the precision and dynamic range compared with those of conventional electrical or photonic sensors. Although this use of a photonic RF sensor has been demonstrated for strain [12], strain/temperature [13,14], and ultrasound [15,16] by use of multiple-longitudinal-mode or multiple-polarization-mode oscillation in a CW fibre laser, use of an OFC will benefit from the ultra-narrow linewidth and high stability of $f_{rep}$ due to the inherent nature of the mode-locking oscillation. However, when a cavity single-mode fibre (SMF) itself is used as a sensor, the measurable sensing quantities are limited to physical quantities that directly interact with the fibre OFC cavity, such as strain [17], acoustic pressure [18], or ultrasound waves [19].

Photonic RF sensors that can sense a wide variety of physical quantities will become possible if a functional fibre sensor can be introduced into a fibre OFC cavity. Such intracavity fibre sensors will benefit from the sensitivity enhancement made possible by the multiple passages of light through the sensor in addition to photonic RF conversion. One interesting application of photonic RF sensors is in sensing of refractive index (RI) because fibre RI sensors have found many applications in quality control of ethanol sensing [20], glucose sensing [21], bio-sensing [22], and gas sensing [23]; however, these applications still need even greater sensing performance.

In this article, we focus on a multi-mode interference (MMI) fibre sensor for RI measurement [24-26]. An MMI fibre sensor is composed of a clad-less multi-mode fibre (MMF) with a pair of SMFs at the two ends, has good compatibility with fibre OFC cavities, and works as an RI sensor based on a change in MMI wavelength $\lambda_{MMI}$ due to the Goos-Hänchen shift on the surface of the clad-less MMF (Fig. 1a, see Methods). An RI-dependent shift of $\lambda_{MMI}$ in the fibre OFC cavity is converted into a shift of $f_{rep}$ via wavelength dispersion of the cavity fibre (Fig. 1b). In our setup, we introduced an MMI fibre sensor into a fibre OFC cavity, forming what we call an MMI-OFC, and read out the change in RI in a sample solution via $f_{rep}$ to demonstrate the potential of MMI-OFCs as photonic RF sensors for RI.

## 2. Results

Figure 2 illustrates the experimental setup of the MMI-OFC, which is described in the Methods section together with details of the experimental and analytical methodologies employed for the following measurements.

Basic characteristics of fibre OFC

Stable mode-locking oscillation in the MMI-OFC is indispensable for high-precision RI sensing. Here we compare the basic characteristics of the MMI-OFC with those of a usual OFC. The blue line in Fig. 3a shows an optical spectrum of a usual OFC comb (centre wavelength = 1561.8 nm, spectral bandwidth = 15 nm, mean power = 4.7 mW), indicating the symmetrical broad spectrum, together with some spikes due to the soliton mode-locking oscillation near the zero-dispersion region of the cavity (-0.045 pm/s$^2$). On the other hand, the red line in Fig. 3a shows an optical spectrum of an MMI-OFC. The spectral bandpass-filtering effect of the intracavity MMI fibre sensor decreased the spectral bandwidth and the mean power to 10 nm and 2.9 mW, respectively. In this way, we could obtain stable mode-locking oscillation even in the MMI-OFC.

Next, we compare the fluctuation and stability of $f_{rep}$ between the MMI-OFC and the usual OFC. $f_{rep}$ in the MMI-OFC was 43.19 MHz, whereas that in the usual OFC was 43.06 MHz. Figure 3b shows a comparison of the fluctuation in $f_{rep}$ between the usual OFC (blue plots) and the MMI-OFC (red plots) with respect to gate time. The fluctuation in $f_{rep}$ is indicated by the standard deviation. The fluctuation in $f_{rep}$ showed similar behaviour in the usual OFC and the MMI-OFC: the fluctuation linearly decreased at gate times below 0.1 s, whereas it increased at gate times over 0.1 s. The drift behaviour of $f_{rep}$ still remained even at gate times greater than 0.1 s, although the temperature of the OFC cavity was stabilized; this behaviour is typical in $f_{rep}$-unstabilized OFCs [27]. $f_{rep}$ in the MMI-OFC reached a minimum value of 0.0265 Hz at a gate time of 0.1 s. Most importantly, the introduction of the MMI fibre sensor into the fibre OFC cavity had little effect on the fluctuation in $f_{rep}$, enabling us to perform high-precision RI sensing based on the stable $f_{rep}$.

RI-dependent shift of optical spectrum

Next, we investigated the shift of the optical spectrum with respect to the sample RI. Ethanol/water solutions with different mixture ratios (= 0–15 EtOH%, corresponding to 1.333–1.340 refractive index units, or RIU) were used as samples with different RIs. We repeated the same experiment for 5 sets of ethanol/water samples with different RIs, and then calculated the mean and the standard deviation of them for each RI. Figure 4a shows a comparison of typical optical spectra of the ethanol/water samples with different RIs. One can confirm the long-wavelength shift of the optical spectrum with increasing RI. Figure 4b show the relation between the sample RI and the shift of the centre wavelength in the optical spectrum. The corresponding ethanol concentration is shown on the upper horizontal axis in Fig. 4b. A positive linear relation between them was confirmed with a slope coefficient of 65.7 nm/RIU, indicated by a blue line. For comparison, we conducted a similar experiment using the same sample by placing the same MMI fibre sensor outside the cavity to use the MMI fibre sensor in the usual way (data is not shown) [26]. The resulting slope coefficient was 73.1 nm/RIU. Therefore, the MMI fibre sensor can work as an RI sensor even inside the fibre OFC cavity, in the same manner as the extracavity MMI fibre sensor.

RI sensing based on $f_{rep}$ shift

Finally, we performed RI measurement of ethanol/water samples with different mixture ratios (= 0–15 EtOH%, corresponding to 1.333–1.340 RIU) based on a shift in $f_{rep}$. Figure 5a shows the RI-dependent shift of the RF spectrum for $f_{rep}$, acquired by an RF spectrum analyser. The linewidth of the RF spectra was limited by the instrumentation resolution of the RF spectrum analyser (= 1Hz) rather than the actual linewidth of the $f_{rep}$ signal. Nevertheless, the amount of spectral shift was significantly larger than the spectral linewidth, compared with the RI-dependent shift of the optical spectrum in Fig. 4a. Such a high ratio of the spectral shift to the spectral linewidth enables high-resolution RI sensing based on RF measurement. Next, we measured $f_{rep}$ values for 5 sets of ethanol/water samples with different RIs by using an RF frequency counter, and we calculated the mean and the standard deviation of them for each RI. Figure 5b shows the relation between the sample RI and the shift in $f_{rep}$ in the MMI-OFC, indicating a negative linear relation between them with a slope coefficient of -6.19×10$^3$ Hz/RIU. Considering the positive relation in Fig. 4b and the negative net dispersion of the cavity fibre in this wavelength band, this negative slope is valid. Since the $f_{rep}$ fluctuation was 0.0265 Hz at a gate time of 0.1 s (see Fig. 3b), the RI resolution was estimated to be $4.28 \times 10^{-6}$. On the other hand, the RI accuracy was estimated to be $5.35 \times 10^{-5}$ when it was defined as the root mean square (RMS) between the experimental data and the linear approximation. The differences of the RI resolution and accuracy are mainly due to the influence of the sample temperature, as discussed later.

## 3. Discussion

We first discuss the validity of the RI-dependent $f_{rep}$ shift (Fig. 5b). An RI-dependent optical spectrum shift is converted into an RI-dependent $f_{rep}$ shift via the wavelength dispersion of the cavity fibre (see Fig. 1b). The present MMI-OFC cavity includes a 2.9-metre SMF with anomalous dispersion (= 17 ps/km/nm) and a 1.6-metre EDF with normal dispersion (= -15 ps/km/nm), leading to a net dispersion of 0.0253 ps/nm. From the slope coefficient of 65.7 nm/RIU in Fig. 4b, the relation between the time delay and sample RI is given as 1.66 ps/RIU, corresponding to $-3.1 \times 10^3$ Hz/RIU. This estimated slope is in reasonable agreement with the experimental plots for the MMI-OFC (= -6,189 Hz/RIU, see Fig. 5b). There are two reasons for the difference between the experimental slope and the estimated one: one is the change in the optical cavity length caused by the Goos-Hänchen shift in the intracavity MMI fibre sensor; the other is the wavelength dispersion of the intracavity MMI fibre sensor. Even including such an influence, the intracavity MMI fibre sensor functions as an RI-dependent tunable bandpass filter without any influence of $f_{rep}$.

Next we discuss the influence of temperature fluctuation in the MMI-OFC cavity because such temporal fluctuation results in thermal expansion or shrinkage of the optical cavity length. Such thermal expansion or shrinkage of the cavity leads to a change in $nL$ and hence $f_{rep}$. We measured the $f_{rep}$ shift in the MMI-OFC with a pure water sample (= 0 EtOH%, RI = 1.333) when the cavity temperature $T_{cav}$ was changed within the range of 20.0 to 21.4 °C. Figure 6a shows the relation between the cavity temperature and $f_{rep}$ shift. A linear relation was confirmed between them, and its slope was determined to be -177 Hz/°C by a linear approximation. On the other hand, the $f_{rep}$ fluctuation in Fig. 3b was 0.0265 Hz at a gate time of 0.1 s. Therefore, if the $f_{rep}$ fluctuation is mainly due to the fluctuation of $T_{cav}$, $T_{cav}$ is estimated to be stabilized within $1.5 \times 10^{-4}$ °C during a time period of 0.1 s.

We also investigated the relation between the sample temperature $T_{sam}$ and the $f_{rep}$ shift because the sample RI depends on both EtOH % and temperature. To this end, we measured the change in $f_{rep}$ when changing $T_{sam}$ of a pure water sample (= 0 EtOH%, RI = 1.333) within the range of 22.0 °C to 25.0 °C, as shown in Fig. 6b. A linear dependence of $f_{rep}$ on $T_{sam}$ was confirmed again; however, the slope constant (= -30.9 Hz/°C) was 6-times smaller than that of $T_{cav}$. Therefore, $T_{cav}$ control is more important than $T_{sam}$ control for high-resolution RI sensing. If the $f_{rep}$ fluctuation is influenced by the fluctuation of $T_{sam}$ rather than that of $T_{cav}$, $T_{sam}$ is estimated to be stabilized within $8.5×10^{-4}$ ºC during a time period of 0.1 s. $T_{sam}$ may further influence the RI accuracy because the reproducibility of $T_{sam}$ in repetitive measurements appears in the RI accuracy. From the RI accuracy of $5.35×10^{-5}$ in Fig. 5b, the reproducibility of $T_{sam}$ is estimated to be within 0.01 ºC. This value is reasonable considering the performance of the temperature controller.

In summary, we integrated an MMI fibre sensor into a fibre OFC for high-precision RI measurement. The RI change of a liquid sample was transferred to a change in $f_{rep}$ via multi-mode interference in the MMI fibre sensor and wavelength dispersion in the cavity fibre. Combined use of a narrow-linewidth, stable RF comb mode with high-precision frequency measurement achieved an RI resolution of $4.28×10^{-6}$ and an RI accuracy of $5.35×10^{-5}$. The concept of a sensing RF comb will expand the application scope of OFCs beyond their current use in optical frequency rulers and will extend to fibre sensing, including RI sensing.

**Methods**

**MMI fibre sensor**

$\lambda_{MMI}$ in the MMI fibre sensor is given by

$$\lambda_{MMI} = \frac{n_{MMF} m}{L_{MMF}} D(n_{sam})^2 \qquad (2)$$

where $n_{MMF}$ and $L_{MMF}$ are, respectively, the refractive index and the geometrical length of the clad-less MMF, $m$ is the order of the MMI, $n_{sam}$ is the sample RI, and $D(n_{sam})$ is the effective diameter of the clad-less MMF [24-26]. Since $D(n_{sam})$ depends on the sample RI via the Goos-Hänchen shift in total internal reflection, $\lambda_{MMI}$ depends on the sample RI. For example, an increase of the sample RI leads to an increase of the Goos-Hänchen shift and the corresponding $D(n_{sam})$, and hence produces a $\lambda_{MMI}$ long-wavelength shift in proportion to the square of $D(n_{sam})$.

A schematic diagram of the MMI fibre sensor is shown in Fig. 1a. The MMI fibre sensor was composed of a clad-less MMF (core diameter = 125 µm, fibre length = 58 mm) with a pair of SMFs at both ends (core diameter= 6 µm, clad diameter = 125 µm, fibre length = 54 mm), in which the diameter of the SMF clad was equal to that of the clad-less MMF core. We set $m$ to 4 because the MMI signal appears as a spectral peak at a wavelength of $\lambda_{MMI}$. Broadband light passing through the input SMF is diffracted at the entrance face of the clad-less MMF, and then repeats total internal reflection at the boundary between the clad-less MMF surface and the sample solution. $\lambda_{MMI}$ light mainly exits through the clad-less MMF and then goes toward the output

SMF. Due to the spectral bandpass-filtering effect in the MMI process, the MMI fibre sensor acts as an RI sensor showing an RI-dependent $\lambda_{MMI}$ shift. We designed the parameters of the MMI fibre sensor so that the spectral peak appeared around the centre wavelength of fibre OFC (= 1555 nm) when the sample is a pure water sample (0 EtOH%, RI = 1.333).

**Experimental setup**

We used a mode-locked Er:fibre laser oscillator for the MMI-OFC (see Fig. 2). This oscillator had a ring cavity including a 2.9 m length of single-mode fibre (SMF, SMF-28, Corning, dispersion at 1550 nm = 17 ps/km/nm), a 1.6 m length of erbium-doped fibre (EDF, ER30-4/125, LIEKKI, dispersion at 1550 nm = -15 ps/km/nm), a polarization controller [PC, PCUA-15-S/F(15P/15Q/15H), Optoquest Co., Ltd.], a polarization-insensitive isolator (ISO, PSSI-55-P-I-N-B-I, AFR), a 70:30 fibre coupler (FC, SBC-1-55-30-1-B-1, AFR), a wavelength-division-multiplexing coupler (WDM, WDM-1-9855-1-L-1-F, AFR), and a pumped laser diode (LD, BL976-PAG900, Thorlabs, wavelength = 980 nm, power = 900 mW). The temperature of the fibre cavity was controlled to 25.0 ºC by a combination of a Peltier heater (TEC1-12708, Kaito Denshi, power = 76 W), a thermistor (PB7-42H-K1, Yamaki), and a temperature controller (TED200, Thorlabs, PID control). Stable, reproducible soliton mode-locking oscillation was achieved by nonlinear polarization rotation with near zero cavity dispersion (= -0.045 pm/s$^2$) before and after introducing the MMI fibre sensor in the cavity. The output light from the oscillator was detected by a photodetector (PD), and its $f_{rep}$ was measured by an RF frequency counter (53230A, Keysight Technologies, frequency resolution = 12 digit/s) and an RF spectrum analyser (E4402B, Keysight Technologies, frequency resolution = 1 Hz), both of which were synchronized to a rubidium frequency standard (FS725, Stanford Research Systems, accuracy = $5\times10^{-11}$ and instability = $2\times10^{-11}$ at 1 s). Also, its optical spectrum was measured by an optical spectrum analyser (AQ6315A, Yokogawa Electric Corp., wavelength accuracy = 0.05 nm, wavelength resolution = 0.05 nm).

**Samples**

Mixtures of ethanol and pure water were used as standard samples. The sample RI was adjusted by changing the mixture ratio between ethanol and water. The relationship between the ethanol concentration $EC$ (EtOH%) and the sample RI (RIU) is given by [28]

$$RI = 1.3326 + 4.90\times10^{-4}\times EC \tag{3}$$

The temperature of the sample was controlled to 22 ºC by a combination of a K-type thermocouple (TJA-550K, AS ONE), a cord heater (603-60-69-01, Tokyo Glass Kikai, power = 15 W), and a temperature controller (TJA-550, AS ONE, PID control, display resolution = 0.1 ºC).

**Data availability**

The data that support the findings of this study are available from the corresponding author upon reasonable request.

## Acknowledgements

This work was supported by grants for the Exploratory Research for Advanced Technology (ERATO) MINOSHIMA Intelligent Optical Synthesizer (IOS) Project (JPMJER1304) from the Japanese Science and Technology Agency.

## Author contributions

T. Y. and S. T. conceived the project. R. O. and K. N. constructed the MMI-OFC, performed the experiments, and/or analysed the data. S. T. and H. F. contributed to design and construction of the MMI fibre sensor. T. M., Y. N., and K. M. contributed to design and improvement of the fibre OFC cavity. R. O. and T. Y. wrote the manuscript. S. T., T. M., Y. M., T. I., H. Y., and T. Y. discussed the results and commented on the manuscript.

## Competing financial interests statement

The authors declare no competing financial interests.

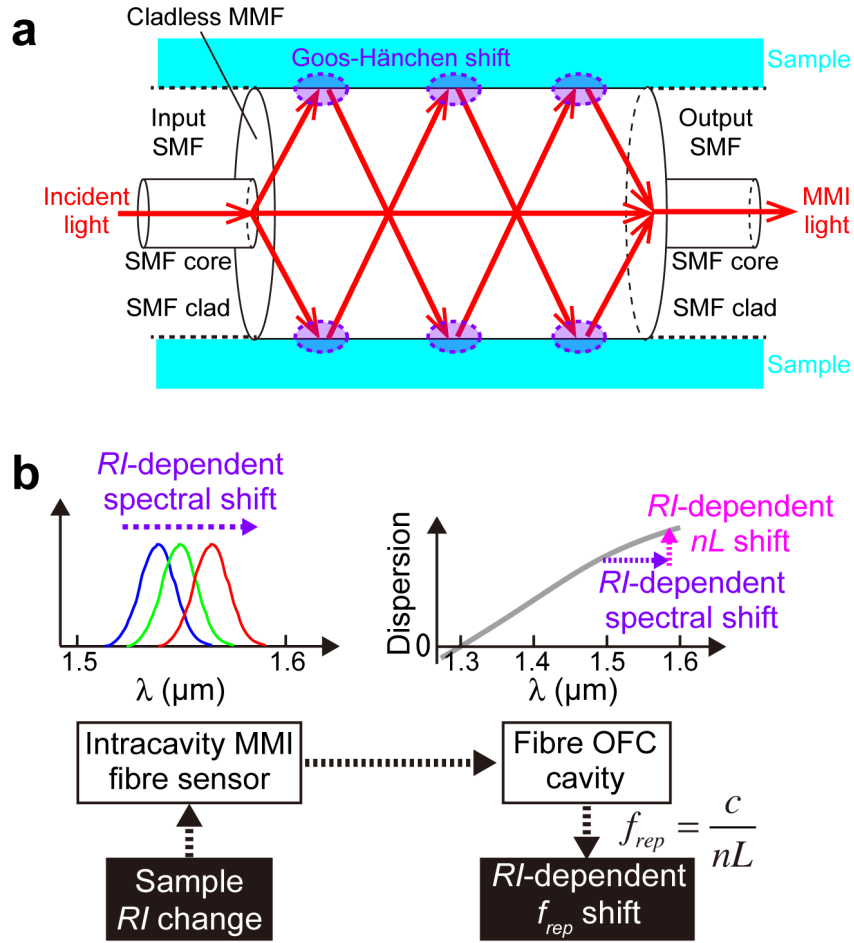

**Figure 1. Principle of operation. a**, Schematic diagram of the MMI fibre sensor. The MMI fibre sensor functions as an RI-dependent tunable bandpass filter via the MMI process. See Methods for details. **b**, Conversion from sample RI change to RI-dependent $f_{rep}$ shift. The intracavity MMI fibre sensor shifts the optical spectrum of the MMI-OFC depending on the sample RI. The wavelength-shifted MMI-OFC spectrum experiences the wavelength dispersion of the cavity fibre, resulting in the conversion from an RI-dependent spectral shift to an RI-dependent shift of the optical cavity length $nL$. Such an RI-dependent $nL$ shift leads to an RI-dependent $f_{rep}$ shift based on Eq. 1.

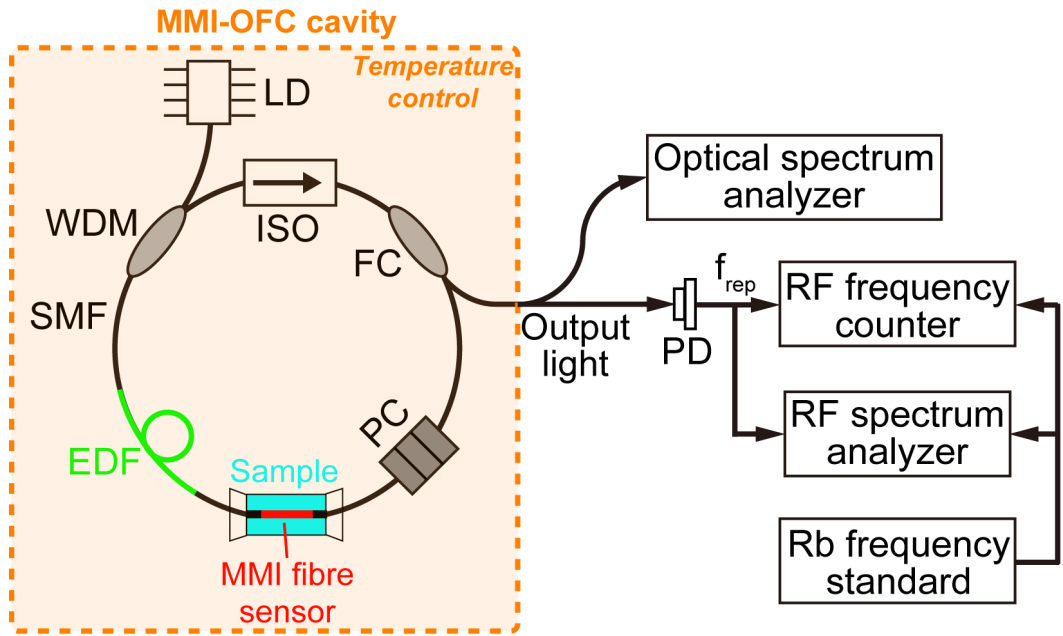

**Figure 2. Experimental setup.** See Methods for details.

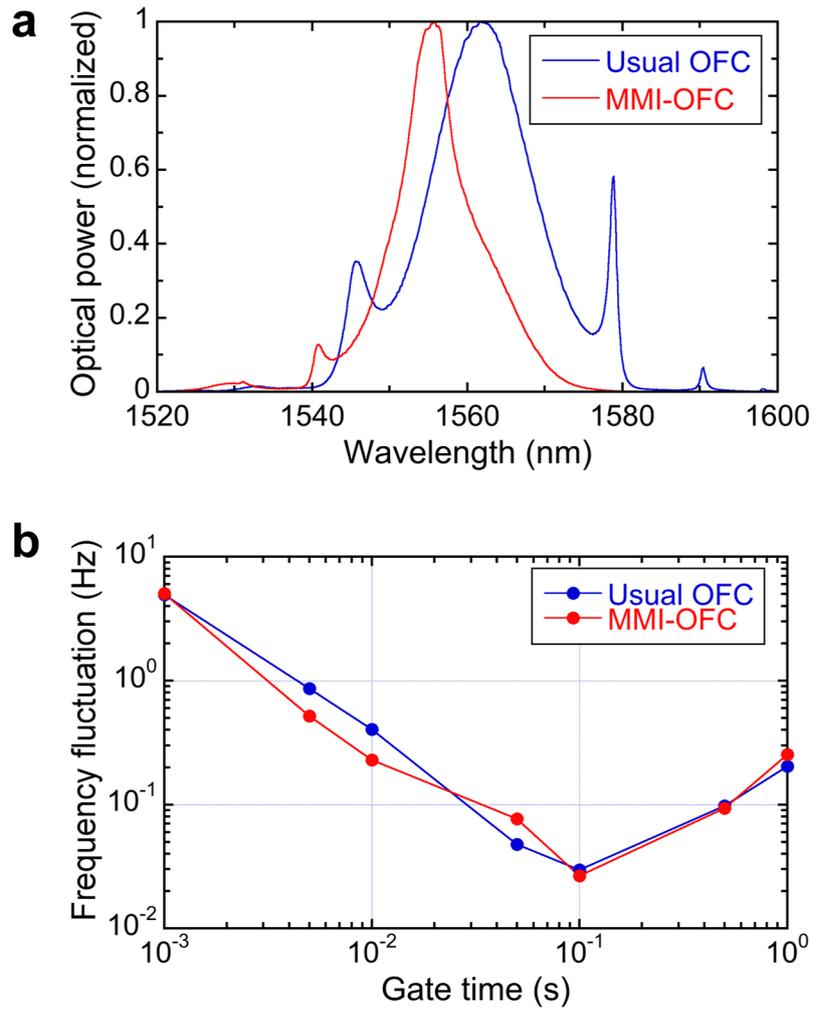

**Figure 3. Basic performance. a**, Comparison of optical spectrum between usual OFC and MMI-OFC. **a**, Comparison of frequency instability between usual OFC and MMI-OFC.

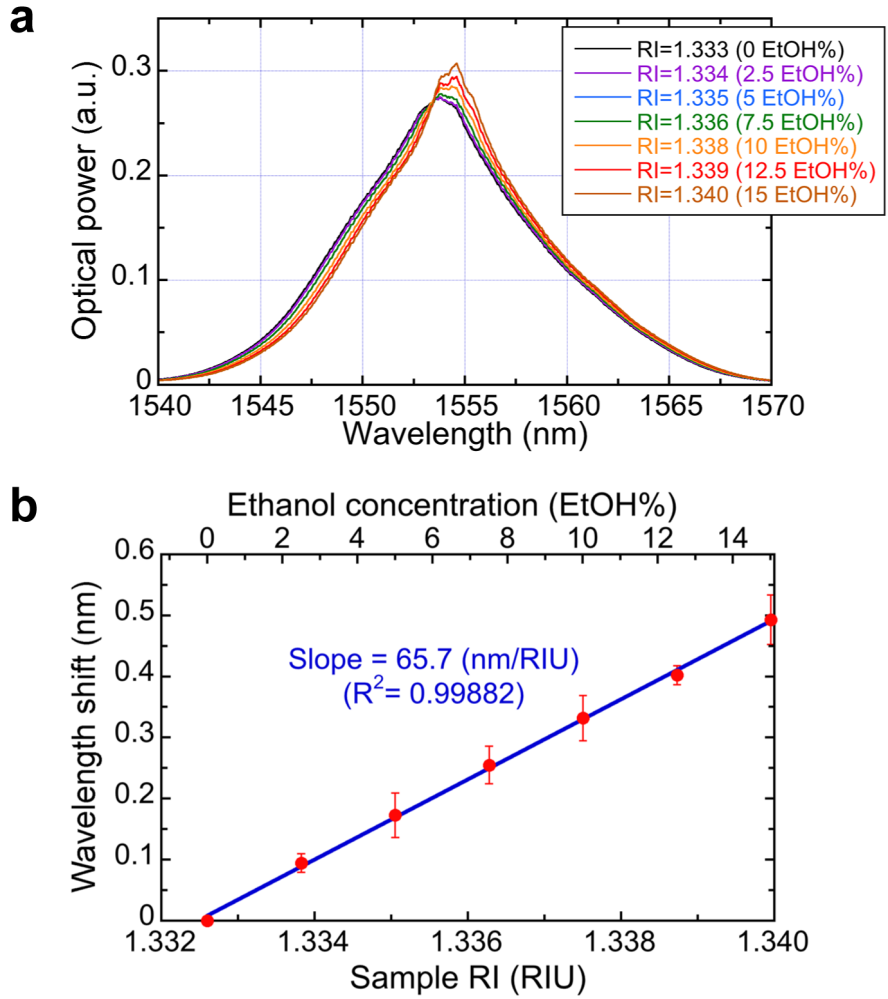

**Figure 4. RI-dependent shift of optical spectrum. a**, Optical spectra of MMI-OFC with respect to different sample RIs. Increasing sample RI causes a long-wavelength shift of the optical spectrum. **b**, Relation between sample RI and wavelength shift $\Delta\lambda$. Plots and error bars indicate the mean and the standard deviation of $\Delta\lambda$ in 5 repetitive measurements. Blue line shows a linear approximation by a curve fitting analysis.

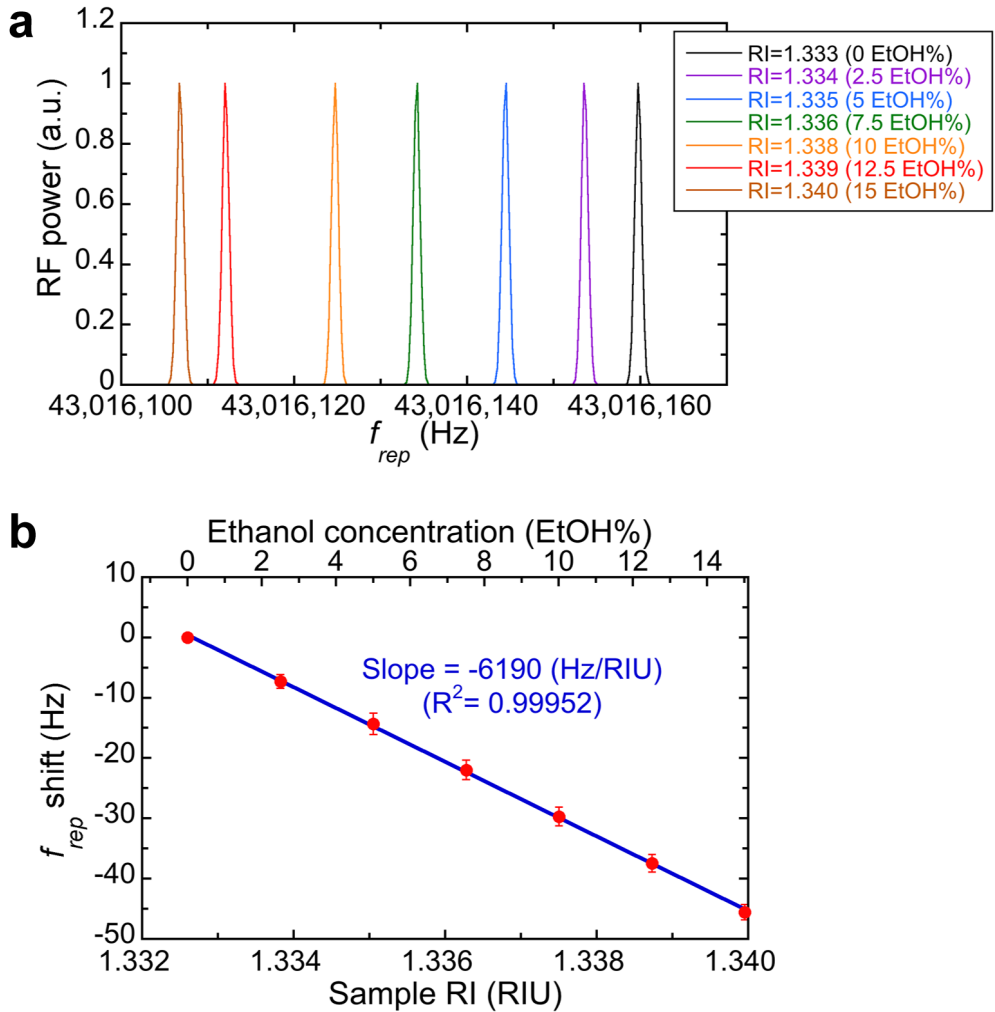

**Figure 5. RI-dependent $f_{rep}$ shift. a**, RF spectra of $f_{rep}$ signal with respect to different sample RI. Increasing sample RI causes decrease of $f_{rep}$. **b**, Relation between sample RI and $f_{rep}$ shift. Plots and error bars indicate the mean and the standard deviation of $f_{rep}$ in 5 repetitive measurements. Blue line shows a linear approximation by a curve fitting analysis.

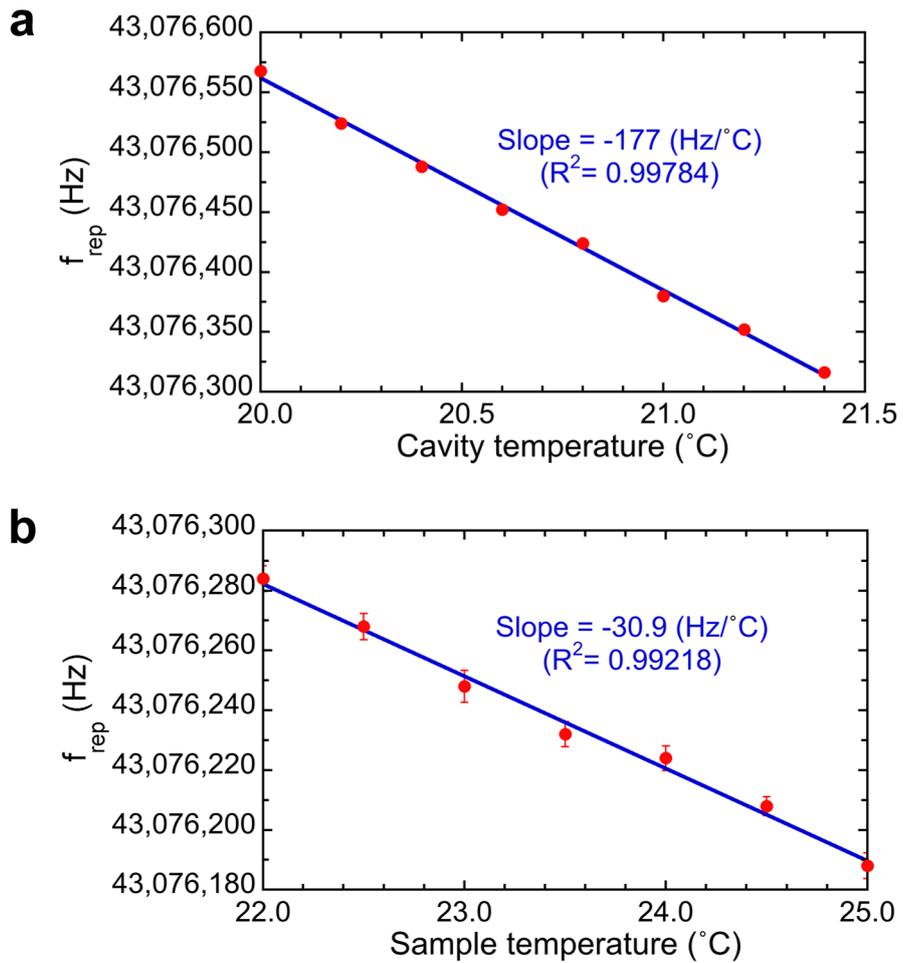

**Figure 6. Dependence of $f_{rep}$ shift on temperature. a**, Cavity temperature dependence. **b**, Sample temperature dependence. Plots and error bars indicate the mean and the standard deviation of $f_{rep}$ in 5 repetitive measurements. Blue line shows a linear approximation by a curve fitting analysis.